\documentclass[pra,aps,twocolumn,superscriptaddress,showpacs,tightenlines,amsmath]{revtex4}

\usepackage{epsfig}

\begin{document}
\title{Storing Images in Entangled Quantum Systems}
\author{S. E. Venegas-Andraca}
\affiliation{Centre for Quantum Computation, Clarendon Laboratory,
University of Oxford, Oxford OX1 3PU, United Kingdom}
\author{J. L. Ball}
\affiliation{Centre for Quantum Computation, Clarendon Laboratory,
University of Oxford, Oxford OX1 3PU, United Kingdom}

\begin{abstract}
We introduce a new method of storing visual information in Quantum
Mechanical systems which has certain advantages over more
restricted classical memory devices. To do this we employ uniquely
Quantum Mechanical properties such as Entanglement in order to
store information concerning the position and shape of simple
objects.
\end{abstract}

\maketitle

\section{Introduction}
The  storage, processing and  retrieval of visual information are
first order tasks for researchers in the discipline of Image
Processing and related areas such as  Pattern Recognition and
Artificial Intelligence. However, due to the restricted
architecture of classical computers and the often overwhelming
computational complexity of state-of-the-art algorithms, it is
necessary to find better ways to store, process and retrieve
visual information.

%Since only a finite number of data values can be transmitted to a
%computer, spatial quantization of any image is required.
%In what follows, we consider an image to be constructed from
%a countable number of pixels.

In a classical memory device, memory cells are, in terms of
hardware, independent of one another, that is, each storage
location can be ascribed a reality that is independent of all
other locations in the memory. Furthermore, such independence is
actually inherited by any kind of information stored in such
memory cells. The only way to correlate values stored in a
classical computer memory is by means of software.

Thus, storing an image in a classical Von Neumann computer involves a memory
which essentially consists of a large set of {\it independent}
bits, each of which represents some property of the associated
image, for example light intensity. Recovery of information
concerning the image involves reading binary data stored in
the computer memory, that is, an image is recovered by
{\it performing independent measurements of a physical property},
that of electric potential difference, on each cell of the memory device.
However, correlations between different points in an
image are very important in order to properly understand and
describe it. Typically, any one part of an image  bears an
important relation to other parts. Since the bits that are used to
store such images in a classical Von Neumann computer are essentially
independent of one another, much relevant information pertaining
to the image may be lost upon classical data storage.
Alternative methods for data storage and processing have been proposed
in the past in order to overcome such loss of information. Among them,
associative memories stand in first place \cite{kohonen}. However,
associative memories is an inefficient proposal as the number of
patterns that can be stored in a $n$-bit associative memory is
$O(n)$. A recent development by Trugenberger \cite{trugenberger}
shows that it is possible to use a multi-particle quantum system
to implement an efficient quantum associative memory.

Methods for data storage, processing and retrieval form a broad
and  active field of research in Quantum Information Processing.
In the next section we review some of the more general principles
of Quantum Information before specializing in later sections to
the subject of image storage and retrieval.

\section{Quantum vs. Classical Computation}

We shall devote this section to an analysis of some fundamental
elements of Quantum Mechanics used in Quantum Computation (QC) and
Quantum Information Processing (QIP).

We start by exploring the differences between the basic
information storage components of classical and quantum computers,
namely bits and qubits. This topic is important due to the
non-trivial physical structure of a qubit and its corresponding
mathematical representation. We also study in this section how
sets of qubits interact and combine, as well reviewing  the
mathematical representation of a quantum system composed of $n$
qubits.

Following this we briefly explain the theory of measurement in
Quantum Mechanics. Measuring in quantum mechanics is far from
being either intuitive or trivial. In fact, measurement  is one of
the most striking features of quantum theory and the concept of
measurement strategy plays a most important role in information
extraction in QC and QIP.

Finally, we look at a unique property of quantum systems that has
no equivalent in the world of classical physics: quantum
Entanglement. Entanglement is a unique type of correlation shared
between components of a quantum system. Entangled quantum systems
are often best used collectively, that is, an optimal use of
entangled quantum systems for information storage and retrieval
must manipulate and measure those systems as a whole, rather than
on an individual basis.

Entanglement has emerged as a key concept in QC and QIP as it is
used as a physical resource to build quantum algorithms
\cite{shor,grover} as well as to develop schemes for quantum
teleportation \cite{bennett}.

\subsection{Classical bits vs. Qubits}

{\bf Mathematical representations of a classical bit and a single qubit}
\\
\\
In recent years, progress in the field of Quantum Computation and
Quantum Information has taught computer scientists that nature can
be harnessed far more efficiently for computation by exploiting
quantum-mechanical properties of systems. In 1985 Deutsch
developed a theoretical machine, the Universal Quantum Turing
Machine, which is a generalization of the Universal Turing Machine
\cite{deutsch}. Such a Quantum Computer, which performs
computations according to the rules of Quantum Mechanics, is
capable of performing certain tasks more efficiently than its
classical counterpart.  Two celebrated examples of such tasks are
the factorization of large numbers in Shor's algorithm \cite{shor}
and the search for a data item in an unordered database in
Grover's algorithm \cite{grover}.

In Classical Computation, information is stored and manipulated in
the form of bits. The mathematical structure of a classical bit is
rather simple. It suffices to define two \lq logical' values,
traditionally labelled as $\{0,1\}$, and to relate these values to
two different outcomes of a classical measurement. So a classical
bit \lq lives' in a scalar space.

In Quantum Computation, information is stored, manipulated and
measured in the form of qubits.  A qubit is a physical entity
described by the laws of Quantum Mechanics. Simple  examples of
qubits include two orthogonal  polarizations of a photon (e.g.
horizontal and vertical), the alignment of a (spin-1/2) nuclear
spin in a magnetic field or  two states of an electron orbiting an
atom. A qubit may be mathematically represented as a vector
$|\Psi\rangle$ in a two-dimensional complex vector space which has
an associated inner product, so $ |\Psi\rangle \in {\cal H}^{2}$.
For the sake of this discussion, we refer to such a vector space
as a two-dimensional Hilbert space ${\cal H}^{2}$.

The notation  $|\rangle$, a ket, is part of the {\it Dirac
notation}, a standard and very convenient typography in Quantum
Mechanics which actually is far more than mere notation.

A qubit $|\Psi\rangle$ may be written in general form as
\begin{equation}
|\Psi\rangle=\alpha|p\rangle+\beta|q\rangle
\end{equation}
where the complex coefficients $\alpha$ and $\beta$ satisfy the
normalization condition $|\alpha|^2+|\beta|^2=1$ and
$\{|p\rangle,|q\rangle\}$ is an arbitrary basis spanning ${\cal
H}^{2}$. The choice of $\{|p\rangle, |q\rangle\}$ is often
$\{|0\rangle,|1\rangle\}$. These are the computational basis
states and form an orthonormal basis for the qubit vector space.
So in general $|\Psi\rangle$ is a coherent superposition of the
basis states $|p\rangle$ and $|q\rangle$ and can be prepared in an
infinite number of ways simply by varying the values of the
complex coefficients $\alpha$ and $\beta$ subject to the
normalization constraint. In contrast, classical computers measure
bit values using only one basis, $\{0,1\}$, and the only two
possible states are those that correspond to the measurement
outcomes 0 or 1.

A qubit can also be represented by a {\it density operator} (often
the {\it density matrix} is used in the literature). Both
representations are equivalent, thus using one representation or
the other depends on the properties of the system to be studied.
The density operator of a qubit is usually denoted as
$\hat{\varrho}$.

For example, it is a good idea to use a vector representation in
problems where we know with certainty the initial state of the
qubit. An example of this statement is to have a qubit prepared in
the state $|\Psi\rangle = \frac{|0\rangle+|1\rangle}{\sqrt{2}}$,
that is, an equally weighted superposition of the canonical basis
$\{|0\rangle, |1\rangle\}$.

However, let us consider a different scenario in which a qubit
$|\Psi\rangle$ is initially prepared in one of the following
quantum states: $\{|\psi\rangle_1, |\psi\rangle_2, |\psi\rangle_3,
\ldots, |\psi\rangle_n \}$ where each of the states is selected
with probability $\frac{1}{n}$. We do not know what state was
chosen to prepare $|\Psi\rangle$, but we do know that only
preparations $|\psi\rangle_i$, $i \in \{1, 2, \ldots, n\}$ are
allowed. In this case, a convenient representation for
$|\Psi\rangle$ is the associated density operator

\begin{equation}
\hat{\varrho}_{\Psi} = {1 \over n} \sum\limits_{k=1}^{n}
|\psi\rangle_k{}_k\langle\psi|
\end{equation}

The symbol $\langle |$ denotes a bra, another symbol from Dirac
notation. $\langle\psi|$ is alternatively written as
$|\psi\rangle^\dagger$, the complex conjugate transpose of
$|\psi\rangle$. As an example, let us set $|\psi\rangle
=\begin{pmatrix}0\\1\end{pmatrix}$  for the computational basis
$\{|0\rangle,|1\rangle\}$. Then, $\langle\psi| =
\begin{pmatrix}0&1\end{pmatrix}$. Taking the inner product yields
$\langle\psi|\psi\rangle =
\begin{pmatrix}0&1\end{pmatrix}\begin{pmatrix}0\\1\end{pmatrix} = 1$ i.e. the inner
product of $|\psi\rangle$ with itself is unity.
\\
\\
{\bf Mathematical representation of an array of qubits}
\\
\\
A composite qubit system $|\Phi\rangle$, often also called a qubit
array, is a quantum system made up of several qubits, where each
qubit is associated with a two-dimensional Hilbert space ${\cal
H}^{2}$. The postulates of Quantum Mechanics establish that the
vector space where a composite qubit system lives is the tensor
product of the vector spaces of the component physical systems.
Thus, if component qubits are $|\psi\rangle_1, |\psi\rangle_2,
\ldots, |\psi\rangle_n$ (where each qubit resides in a
two-dimensional Hilbert space ${\cal H}^{2}$) then $|\Phi\rangle$
lives in a Hilbert space ${\cal H}^{2^n} = {\cal H}^{2} \otimes
{\cal H}^{2} \otimes \ldots \otimes {\cal H}^{2}$.

%The composite qubit system
%$\ket{\Phi}$ is an element of a Hilbert space with dimension
%$2^n$.
%$\ket{\psi}_i$
%let Let $\ket{\psi}_i$, $ i \in \{1, 2, \ldots, n\}$
%be $n$ qubits. Then
%\begin{equation}
%\ket{\Phi}= \bigotimes_{i=1}^{i=n} \ket{\psi}_i
%\end{equation}
%is a composite quantum system living in a Hilbert space ${\cal H}^{2^n}$.

A simple classical data storage device can be considered as an
array made up of $n$ classical bits and capable of storing one of
$2^n$ different classical bit strings. A simple quantum memory
device is a qubit array. Such a data storage device consists of
$n$ qubits. Each is prepared in some quantum state determined by
the desired information to be stored in that particular qubit.
Thus, in principle, both classical and quantum arrays can store up
to $2^n$ different values. However, in stark contrast to classical
data storage, an array of $n$ qubits is capable of storing in a
coherent superposition $2^n$ different bit strings {\it
simultaneously} \cite{trugenberger}.

Excellent introductions to vector and density matrix
representations for single and multiple qubits can be found in
\cite{rieffel,nielsen}.

\subsection{Quantum Measurement}

Measurement according to the rules of Quantum Mechanics is a
non-trivial and highly counter-intuitive process. Firstly, it must
be said that the measurement results taken from a quantum system
are inherently of a probabilistic nature. In other words,
regardless of the carefulness in the preparation of a measurement
procedure, the possible outcomes of such measurement will be
distributed according to a certain probability distribution.

Secondly, once a measurement has been performed,  a quantum system
in unavoidably altered due to the interaction with the measurement
apparatus. Thus, it makes sense to talk about pre-measurement and
post-measurement quantum states for an arbitrary quantum system.

Thirdly, in order to perform a measurement it is needed to define
a set of measurement operators. This set of operators must fulfill
a number of rules that allows one  to compute the actual
probability distribution as well as post-measurement quantum
states.

In order to clarify these points, let us work out a simple
example. Assume we have a polarized photon with associated
polarization orientations \lq horizontal' and \lq vertical'. The
horizontal polarization direction is denoted by $|0\rangle$ and
the vertical polarization direction is denoted by $|1\rangle$.

Thus, an arbitrary initial state for our photon can be described
by the state by $|\Psi\rangle =\alpha|0\rangle +\beta|1\rangle$,
where $\alpha$ and $\beta$ are complex numbers constrained by the
normalization condition $|\alpha|^2+|\beta|^2=1$ and $\{|0\rangle,
|1\rangle\}$ is the computational basis spanning ${\cal H}^{2}$.

Now, let us construct two measurement operators $\hat{M}_0 =
|0\rangle\langle0|$ and $\hat{M}_1 = |1\rangle\langle 1|$ and two
measurement outcomes $a_0, a_1$. Then, the full {\it observable}
used for measurement in this experiment is $\hat{M} =
a_0|0\rangle\langle 0| + a_1|1\rangle\langle 1|$.

According to the rules of Quantum Mechanics, the probabilities of
obtaining outcome $a_0$ or outcome $a_1$ are given by $ p(a_0) =
|\alpha|^2$ and $ p(a_1) = |\beta|^2$. Corresponding
post-measurement quantum states are as follows: if outcome $= a_0$
then $|\Psi_{pm}\rangle = |0\rangle$; if outcome $= a_1$ then
$|\Psi_{pm}\rangle = |1\rangle$.

It is possible to construct a full quantum measurement theory for
both vector and density matrix representations of quantum systems
(for a review, see \cite{nielsen}). Measurement theory and its
implications in QC and QIP are open and fruitful fields of
research.

\subsection{Quantum Entanglement}

Quantum Entanglement is a key concept in QC and QIP. We shall
review the experiments that sparked an investigation into quantum
entanglement as well as the mathematical structure of some
entangled states widely used in QC and QIP.

The concept of correlation is deeply rooted in every branch of
science. A typical and simple example is the following experiment:
let us suppose we have two balls, one white and one black, as well
as two boxes. If we randomly put a ball in each box and then close
both boxes, we need to perform only one experiment, that is, to
open one box, in order {\it to know which of the balls is in each
box}. In other words, by means of one measurement, opening one box
and seeing which ball was stored in it, we obtain two pieces of
information, namely the colour of the ball stored in both boxes.

The former experiment is an example of classical correlation.
Quantum entanglement is also a kind of correlation, but one that
is detected only in quantum phenomena.

Quantum systems can be tested for inherent entanglement using
various measurement procedures. During the development of Quantum
Mechanics as a physical theory, it was discovered that it is
either an incomplete theory of nature, or that it is nonlocal,
meaning that something which happens spontaneously at one place
instantaneously influences what is true at another place (for
historical discussion see e.g. \cite{EPR}). In order to prove that
Quantum Mechanics is not an incomplete theory, John Bell developed
a series of inequalities to test the validity of Quantum Mechanics
(for a more detailed discussion see e.g. \cite{bell}). Violation
of the bipartite Bell inequalities implies that some
quantum-mechanical predictions cannot be reproduced by a local
hiddenvariable model.

Before elaborating more on Bell inequalities, let us review  the
mathematical appearance of entangled systems for simple cases.

For example, consider the following 2-particle state:

\begin{equation}
|\Psi_-\rangle =\frac{|01\rangle - |10\rangle}{\sqrt{2}}
\end{equation}
(This is the famous EPR/singlet state which appears in many
discussions of non-locality). Clearly, $|\Psi_-\rangle$ lives in a
four-dimensional Hilbert space. It can be seen, after some
calculations, that it is impossible to find quantum states
$|a\rangle, |b\rangle \in {\cal H}^{2}$ such that $|a\rangle
\otimes |b\rangle = |\Psi\rangle$, that is, $|\Psi\rangle$ is not
a product state of $|a\rangle$ and $|b\rangle$. This is indeed a
criterion to determine whether a quantum state is entangled or
not, whether it is possible to express such a composite quantum
state as a simple tensor product of quantum subsystems.

Another example is the tripartite entangled GHZ state

\begin{equation}
|\text{GHZ}\rangle = \frac{|000\rangle + |111\rangle}{\sqrt{2}}
\end{equation}

Again, it is not possible to find three quantum states $|a\rangle,
|b\rangle, |c\rangle \in {\cal H}^{2}$ such that $|a\rangle
\otimes |b\rangle \otimes |c\rangle = |\text{GHZ}\rangle$.

It must be said that entanglement definition and quantification is
an open research area. Currently it is known how to identify and
quantify entanglement for two particles. For three or more
particles the situation is far less straightforward and remains an
active area of research.

%Another uniquely quantum-mechanical property that has no classical
%counterpart is that of Quantum Entanglement.

Let us now turn to some basic definitions and extensions of the
previous theory in the density matrix formalism. Consider a
bipartite system with Hilbert space ${\cal H}={\cal
H}_1\otimes{\cal H}_2$. Suppose that two qubits reside in the
joint state $|\Psi\rangle_{12}$. This state is said to be
separable if it is possible to write
$|\Psi\rangle_{12}=|\Psi\rangle_1\otimes|\Psi\rangle_2$. Such a
bipartite state may instead be written in terms of its density
matrix $\hat{\varrho}$ ($\hat{\varrho}=|\Psi\rangle\langle\Psi|$
for a pure state $|\Psi\rangle$), in which case any separable
state may be written as a convex sum of direct-product states
\begin{equation}
\hat{\varrho}^{(12)}=\sum_ip_i\hat{\varrho}^{(1)}_i\otimes\hat{\varrho}^{(2)}_i
\label{Eq:sep}
\end{equation}
where the $p_i$ represent probabilities satisfying the condition
$\sum_ip_i=1$.  An entangled state is a state which
\textit{cannot} be written in the form of Eq.~(\ref{Eq:sep})
above. In this case, it is impossible to ascribe an independent
reality to either of the two subsystems separately. Instead, they
share only a collective reality.

The basic concepts of bipartite entanglement may be extended to
the multipartite case.  For example, a tripartite state is
unentangled if it is possible to write the associated density
matrix in the form
\begin{equation}
\hat{\varrho}^{(123)}=\sum_{ijk}p_{ijk}\hat{\varrho}^{(1)}_i\otimes\hat{\varrho}^{(2)}_j\otimes\hat{\varrho}^{(3)}_k
\end{equation}
where
$\hat{\varrho}^{(1)}_i,\hat{\varrho}^{(2)}_j,\hat{\varrho}^{(3)}_k$
represent single-particle density matrices. A partially-entangled
tripartite state may be written in the form
\begin{equation}
\hat{\varrho}^{(123)}=p_r\hat{\varrho}^{(12)}\otimes\hat{\varrho}^{(3)}
+p_s\hat{\varrho}^{(13)}\otimes\hat{\varrho}^{(2)}+p_t\hat{\varrho}^{(23)}\otimes\hat{\varrho}^{(1)}
\label{Eq:partialent}
\end{equation}
where the $\hat{\varrho}^{(ij)}$ represent entangled states of two
of the subsystems involved.  Any state $\hat{\varrho}^{(123)}$
that obeys $\hat{\varrho}^{(123)}\neq\sum_ip_i\hat{\varrho}_i$,
where all the $\hat{\varrho}_i$ are separable into products of
states of less than three parties, is fully entangled.

For an $N$-qubit system, following state can be defined and will
prove useful in what follows:
\begin{equation}
|\Psi_N\rangle=\frac{|0\rangle^{\otimes N}+|1\rangle^{\otimes
N}}{\sqrt{2}}
\end{equation}
This state is often referred to as the $N$-particle GHZ state (see
e.g.\cite{GHZ}). States of this form can be produced to a good
approximation in quantum optical systems. Experimental
realizations are presented in \cite{rausch,sackett,pan}, where it
is discussed that states with $N$ up to 4 have already been
produced.

In certain cases Bell inequalities  may also be used to detect the
presence of entanglement. Bell-type inequalities for $N$-particle
systems have been derived under the assumption of total and
partial separability and provide us with a way of deciding whether
a set of $N$ particles resides in a maximally entangled state (see
e.g. the Seevinck-Svetlichny inequalities
\cite{svetlichny2,gisinbell}). Such inequalities also provide,
upon violation, experimentally accessible conditions for full
$N$-particle entanglement and will prove very useful in what
follows.

%Quantum systems can be tested for inherent entanglement using
%various measurement procedures. During the development of Quantum
%Mechanics as a physical theory, it was discovered that it is
%either an incomplete theory of nature, or that it is nonlocal,
%meaning that something which happens spontaneously at one place
%instantaneously influences what is true at another place (for
%historical discussion see e.g. \cite{EPR}). This essentially means
%that the particles involved share extra information over and above
%that predicted by quantum mechanics. Such extra information is
%described in terms of hidden variables [REFS HERE]. For bipartite
%entangled systems certain Bell inequalities have been derived in
%order to test for hidden variables.  Violation of the bipartite
%Bell inequalities implies that quantum-mechanical predictions
%cannot be reproduced by a local hidden variable model and may also
%be used to indicate the presence of entanglement. Bell-type
%inequalities for $N$-particle systems have been derived under the
%assumption of total and partial separability and  provide us with
%a way of deciding whether a set of $N$ particles resides in a
%maximally entangled state (see e.g.\cite{svetlichny2,gisinbell}).
%Such inequalities also provide, upon violation, experimentally
%accessible conditions for full $N$-particle entanglement and will
%prove very useful in what follows.

\section{New method for storing images}

In the discussion that follows, we focus our attention on simple
binary images  (those images having only two brightness levels,
black and white). Such images can be obtained quite simply by
thresholding any gray-level image.  Whilst restricted in
application, such images are of interest because they are
relatively straightforward to process and therefore provide a
useful starting point for introducing Entanglement in the context
of image processing.
\subsection{Storage of information}

We aim to store information concerning the structure and content
of a simple image in a quantum system.  Consider an array of $n$
qubits which we propose to use as our memory storage. Each qubit
in the array may be associated with two parameters, $x$ and $y$,
which together  represent grid points of some simple 2D image.
Such an array can therefore be used to store visual information.
Prior to inputting information into the array, we suppose that
each qubit is initialized to state $|0\rangle$. The initial state
of the memory is therefore given by the following expression
\begin{equation}
|\Psi_{\text{initial}}\rangle=\bigotimes^n_{i=1}|0\rangle_{i(x,y)}
\end{equation}

We wish to store information about the position and shape of
certain simple objects which are represented on our grid as
collections of  points. Extending the classical binary image
formalism to qubits, we associate a white point on the grid with
qubit state $|0\rangle$, whilst black corresponds to state
$|1\rangle$. However certain extensions of the classical approach
are necessary to fully exploit the unique properties of
Entanglement. A simple example will suffice to explain the
principles of such a quantum storage device.

\begin{figure}[h]
\epsfig{width=2in,file=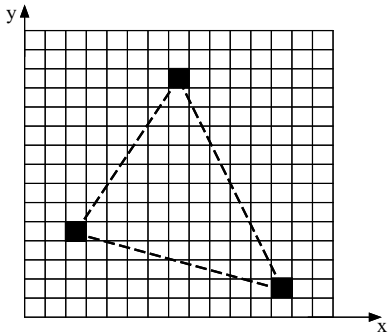} \caption{Simple storage
procedure for a single triangle in a qubit array. In the classical
approach, vertex positions correspond to qubit state $|1\rangle$
and the triangle image can be straightforwardly reconstructed.
However, the use of  Entanglement between vertex locations (dotted
lines) provides a more fruitful approach.} \label{fig:Images}
\end{figure}

Suppose that we wish to store the shape of a triangle in our qubit
array. In this case, we might choose to represent each vertex of
the triangle on the grid by setting the corresponding qubit to
$|1\rangle$. Such a procedure is depicted in
Fig.~(\ref{fig:Images}).

The appropriate vertex positions may then be retrieved by applying
Grover's quantum search algorithm to the array. We would expect
that searching an $n$-qubit array for a $|1\rangle$ using a
classical algorithm would take approximately $O(n)$ steps.
However, Grover's quantum search algorithm can achieve such a task
in approximately $O(\sqrt{n})$ steps due to its use of Quantum
Mechanics. For three vertices stored in the array, application of
Grover's search algorithm will require approximately $\sqrt{n/3}$
steps to recover the information specifying the locations of the
vertices of the triangle. The image of the triangle is then very
simply reconstructed from this information.

\begin{figure}[h]
\epsfig{width=2in,file=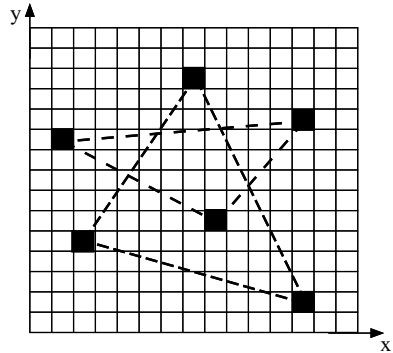} \caption{When two distinct
shapes are stored in the array, entanglement (represented by
dashed lines) between vertices belonging to the same shape is used
to distinguish them from the other.} \label{fig:Images2}
\end{figure}

However, suppose that we instead wish to store {\it two} triangles
in the array. We could proceed as before with an essentially
classical approach, preparing the qubits corresponding to triangle
vertices in state $|1\rangle$ whilst all others remain in state
$|0\rangle$. However, retrieval of information on vertex position
by applying Grover's search algorithm will not reveal anything
about {\it which} vertices belong to {\it which} triangles.  We
need to store additional information in the array concerning which
vertex points belong to which triangle. In this case, Entanglement
may be employed to establish nonlocal correlations between the
qubits storing the vertex locations of the \emph{same} triangle.
Consider again the maximally-entangled tripartite state
\begin{equation}
|\text{GHZ}\rangle=\frac{|000\rangle+|111\rangle}{\sqrt{2}}
\end{equation}
Suppose that our qubit array stores a triangle by preparing the
associated vertex qubits $\{p,q,r\}$ in a GHZ state.  In this
case, the memory state of the qubit array is
\begin{equation}
|\Psi_{\text{1 triangle}}\rangle=\otimes^{n}_{i=1,i\neq
p,q,r}|0\rangle_i\otimes\frac{|000\rangle_{pqr}+|111\rangle_{pqr}}{\sqrt{2}}
\end{equation}
Input of a second triangle with corresponding vertex qubits
$\{s,t,u\}$ into the array yields memory state
\begin{equation}
|\Psi_{\text{2 triangles}}\rangle=\otimes^{n}_{i=1,i\neq
p,q,r,s,t,u}|0\rangle_i\otimes|\text{GHZ}\rangle_{pqr}\otimes|\text{GHZ}\rangle_{stu}
\end{equation}
Retrieval of the information regarding which particles reside in
such maximally entangled states is therefore sufficient to locate
the positions of the triangle vertices and also learn to which
triangle they belong, as depicted in Fig.~(\ref{fig:Images2}).

\subsection{Retrieval}

Once information about an image has been stored in the qubit array
it is desirable to retrieve this information in order to reliably
reconstruct the image. Information retrieval is achieved by way of
performing measurements on the array.

Suppose that we store a triangle in the array in the form of the
state  $|\text{GHZ}\rangle_{pqr}$.  Information pertaining to
relations between one memory location for the image and another
could be retrieved from the array by implementing the measurement
projection operator
\begin{equation}
\hat{M}_{pqr}=|00...0\rangle|\text{GHZ}\rangle_{pqr}{}_{pqr}\langle\text{GHZ}|\langle00...0|
\end{equation}
where $|00...0\rangle$ acts on all qubits not belonging to the GHZ
state. However, since any GHZ state consists of a coherent
superposition of $|000\rangle$ and $|111\rangle$, the qubit array
has the form of a coherent superposition $|\Psi_{\text{1
triangle}}\rangle=(1/\sqrt{2})(\otimes^n_{i=1}|0\rangle+\otimes^n_{i\neq
p,q,r}|0\rangle\otimes|111\rangle_{pqr})=(1/\sqrt{2})(|\Psi_{\text{initial}}\rangle+\bigotimes^n_{i\neq
p,q,r}|0\rangle_i\otimes|111\rangle_{pqr})$. In fact, any memory
state of the qubit array will consist of a coherent superposition
of $|\Psi_{\text{initial}}\rangle$ and other memory states.
Therefore memory states associated with different images are
nonorthogonal and cannot be distinguished unambiguously. This
means that using projection operators will only give probabilistic
results for vertex location and the image cannot be reliably
reconstructed.

Instead, a measurement probing  the \emph{entanglement} shared
between the vertex qubits is employed in order to determine their
location. We illustrate this once again with  our simple triangle
example.

To search for triangles, a set of three qubits in the array is
chosen for measurement. This tripartite state is then tested for
violation of the Seevinck-Svetlichny inequalities for tripartite
states. Violation implies the presence of full three-particle
entanglement. Non-violation therefore implies that the three
qubits selected \textit{do not} form vertices of the same
triangle. From this information it is straightforward to deduce
the location of the triangle vertices. Now suppose that the three
qubits selected consist of two qubits residing in the
\textit{same} GHZ state and a third that does not. Then the state
to be tested is of the form presented in Eq.~(\ref{Eq:partialent})
and will not violate the appropriate inequalities for \emph{full}
tripartite entanglement. For our simple example of two triangles,
determinations of the locations of all six vertices requires at
worst ${}^nC_3.{}^{n-3}C_3$ different identically-prepared arrays
to be tested for two instances of tripartite entanglement amongst
different qubits.

Indeed, suppose that a shape in an image has $N$ vertices. In this
case, an $N$-particle GHZ state is used to store such information.
$N$-particle Bell-type inequalities that provide
experimentally-accessible sufficient conditions for full
$N$-particle entanglement have been derived (see references in
Section 2) and therefore in principle our qubit array may be
tested according to such inequalities to reveal vertex locations
of more complex polygons.

Evidently the construction of a measurement procedure on  the
qubit array requires some \textit{a priori} knowledge of the
number and type of shapes stored in such an array. This
information may be stored in a subset of the array qubits. Such a
subset is addressed first in order to determine the appropriate
number of qubits to pick from the array and test for shared
entanglement, although this is not totally necessary.

\subsection{Use of entanglement for scale-invariant shape
recognition}

\begin{figure}[h]
\epsfig{width=2.5in,file=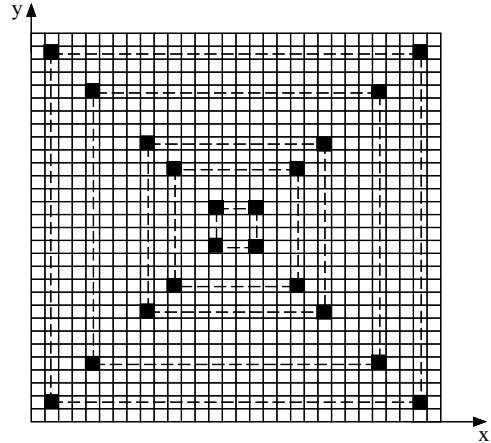} \caption{Simple illustration
of scale-invariant shape recognition using entanglement in a 2D
qubit array. } \label{fig:Images6}
\end{figure}

We briefly note here that Entanglement may also be used to store
and subsequently recognize various shapes in an image irrespective
of their scale. It seems reasonable to suppose that a simple shape
is recognized primarily by the number of vertices it has.  Then
storage of such a shape of \textit{any} size in a qubit array
where entanglement is shared between qubits corresponding to
vertices of the same shape allows it to be recognized irrespective
of its size.  For example, the presence of a 4-particle GHZ state
in a qubit array indicates that the stored image contains a shape
with 4 sides. This information is of course unrelated to the scale
of the shape (see Fig.~(\ref{fig:Images6}). Of course, though, it
is possible to locate the vertex qubits on the 2D grid and deduce
the size of the object quite straightforwardly.

Storage of simple shapes using Entanglement also allows images
that are stored in different memory arrays to be compared by
measurement for similar or identical components simply by
employing the procedures presented in Section IIIB.

\section{Conclusions and Future Work}

We have developed a novel method for storage and retrieval of
simple binary images in Entangled quantum systems.  Our work so
far has concerned only binary images containing simple polygons.
We hope to extend and generalize our work to gray-level images of
increased structural complexity and present this work in a later
paper.
\section*{Acknowledgements}
We thank Sougato Bose and Konrad Banaszek for their support and
encouragement.  This work was supported by EPSRC and CONACyT
(scholarship 148528).

\end{document}